\documentclass[twocolumn, twocolappendix]{aastex631}
\usepackage{bm}% bold math
\usepackage{url}
\usepackage{soul}

\begin{document}

%\title{Scrutinizing Galactic Diffuse Gamma Rays and Neutrinos at TeV-PeV Energies:\\LHAASO-IceCube Connection} 

%\title{Decomposing the Origin of TeV-PeV Neutrinos and Gamma Rays from the Galactic Plane: Implications of the Connection between IceCube and Air Shower Gamma-ray Observations} 

\title{Decomposing the Origin of TeV-PeV Emission from the Galactic Plane: \\ Implications of Multimessenger Observations}

\author[0000-0002-5387-8138]{Ke Fang}
\affiliation{Department of Physics, Wisconsin IceCube Particle Astrophysics Center, University of Wisconsin, Madison, WI, 53706 }

\author[0000-0002-5358-5642]{Kohta Murase}
\affiliation{Department of Physics; Department of Astronomy \& Astrophysics; Center for Multimessenger Astrophysics, Institute for Gravitation and the Cosmos, The Pennsylvania State University, University Park, PA 16802, USA}
\affiliation{School of Natural Sciences, Institute for Advanced Study, Princeton, NJ 08540, USA}
\affiliation{Center for Gravitational Physics and Quantum Information, Yukawa Institute for Theoretical Physics, Kyoto University, Kyoto, Kyoto 606-8502, Japan}

\date{\today}% It is always \today, today,

\begin{abstract}
High-energy neutrino and $\gamma$-ray emission has been observed from the Galactic plane, which may come from individual sources and/or diffuse cosmic rays. We evaluate the contribution of these two components through the multimessenger connection between neutrinos and $\gamma$ rays in hadronic interactions. We derive maximum fluxes of neutrino emission from the Galactic plane using $\gamma$-ray catalogs, including 4FGL, HGPS, 3HWC, and 1LHAASO, and measurements of the Galactic diffuse emission by Tibet AS$\gamma$ and LHAASO. We find that the IceCube Galactic neutrino flux is larger than the contribution from all resolved sources when excluding promising leptonic sources such as pulsars, pulsar wind nebulae, and TeV halos. Our result indicates that the Galactic neutrino emission is likely dominated by the diffuse emission by the cosmic-ray sea and unresolved hadronic $\gamma$-ray sources. 
%In addition, the sum of the flux of non-pulsar sources and the LHAASO diffuse emission in the 1-30~TeV range is compatible with or somewhat larger than the Galactic neutrino flux. 
In addition, the IceCube flux is comparable to the sum of the flux of non-pulsar sources and the LHAASO diffuse emission especially above $\sim 30$~TeV.
This implies that the LHAASO diffuse emission may dominantly originate from hadronic interactions, either as the truly diffuse emission or unresolved hadronic emitters. Future observations of neutrino telescopes and air-shower $\gamma$-ray experiments in the Southern hemisphere are needed to accurately disentangle the source and diffuse emission of the Milky Way. 
\end{abstract}

\section{Introduction}
High-energy neutrinos from the Galactic plane (GP) may come from two components of the Galaxy: the cosmic-ray sea and individual sources. 
The cosmic-ray sea is a smooth and steady distribution of cosmic rays that emerge from accelerators and propagate in the Galactic magnetic field. Protons and nuclei at TeV to PeV energies may be confined in the Galactic magnetic field for 0.1 to a few million years and lose their initial directions. They collide with gas in the interstellar medium (ISM) and produce charged and neutral pions, which decay into neutrinos and $\gamma$ rays, respectively. These secondary particles form the Galactic diffuse emission (GDE). In addition to hadronic cosmic rays, a lower flux of cosmic-ray electrons may also up-scatter the interstellar radiation field and the cosmic microwave background (CMB) to $\gamma$ rays. Above 10~TeV,  electrons have a cooling time of $t_e\sim 64\,(E_e / 10\,\rm TeV)^{-1}\,\rm kyr$ due to the inverse Compton radiation, and propagate for a distance $d \sim (D\,t_e)^{1/2} = 0.3\,(E_e / 10\,\rm TeV)^{-0.33}\,\rm kpc$, where $D \approx 3\times 10^{28}\,(R / 3\,\rm GV)^{1/3}\,\rm cm^2\,s^{-1}$ is the diffusion coefficient assuming Kolmogorov turbulence and $R \equiv E/Ze$ is the rigidity of a particle with energy $E$ and charge number $Z$. Therefore, electrons above tens of TeV cannot travel too far away from the sources where they were produced. 

%Summarize GDE observations by LHAASO, Tibet, HAWC + last generatios (MILAGRO + CASA-MIA) 
GDE in $\gamma$ rays has been measured by the {\it Fermi} Large Area Telescope (LAT) between 100~MeV and 1~TeV over the full sky \citep{2012ApJ...750....3A, 2022ApJS..260...53A}. Above 1~TeV, the GDE from several regions in the Northern sky has been measured by air shower $\gamma$-ray experiments, including ARGO-YBJ at 0.35-2~TeV \citep{2015ApJ...806...20B}, Tibet AS$\gamma$ Observatory at 100-1000~TeV \citep{TibetDiff}, HAWC Observatory at 0.3-100~TeV \citep{HAWC:2021bvb}, and the Large High Altitude Air Shower Observatory (LHAASO) at 10-1000~TeV \citep{2023arXiv230505372C}. 

High-energy neutrinos and $\gamma$ rays may also be produced by individual sources harbored in the Milky Way. About two hundred Galactic $\gamma$-ray sources have been observed above 1~TeV \footnote{http://tevcat.uchicago.edu}. Which sources among them are hadronic emitters, and hence neutrino sources, remains a major question \citep{Sudoh:2022sdk}. One of the challenges arises from the fact that the pion decay and inverse Compton radiation may yield similar spectra. Only a handful sources show promising features of hadronic $\gamma$-ray emission, such as the star formation region at the Galactic center \citep{2016Natur.531..476H} and the supernova remnant G$106.3+2.7$ \citep{2022PhRvL.129g1101F}. To date, no Galactic neutrino sources have been identified. 
%High-energy neutrinos would be a clean messenger to probe the hadronic  channel but 

In addition to resolved sources, unresolved sources may also contribute to emission from the GP. These unresolved sources may be counted toward GDE in measurements despite that they do not have a diffuse nature. The luminosity function of TeV sources is poorly known due to the limited number of sources and the complications related to TeV catalog creations. Based on 32 sources with flux above 10\% Crab from the H.E.S.S. Galactic plane survey (HGPS), the cumulative $\log N-\log S$ distribution of integral flux above 1~TeV is derived to follow a power law with a slope  of $-1.3\pm 0.2$ \citep{HESS:2018pbp}. The distribution is flatter below 10\% although the measurement is limited by the completeness of the sample. Based on the luminosity function derived from the HGPS sources \citep{HESS:2018pbp, 2020A&A...643A.137S}, it has been suggested that the GDE flux measured by Tibet AS$\gamma$ and LHAASO may come from unresolved pulsar-powered sources that are presumably leptonic \citep{Cataldo_2020, 2022ApJ...928...19V, 2023arXiv230506948Z}.

The population of TeV sources has grown significantly following the launches of air shower detectors. The Third HAWC Catalog (3HWC) reported 65 sources, including 20 sources that are more than $1^\circ$ away from any previously detected TeV source \citep{2020ApJ...905...76A}. The first LHAASO catalog (1LHAASO) reported 90 sources, out of which 43 are detected above 100~TeV at $>4\,\sigma$ \citep{2023arXiv230517030C}. However, the luminosity function has been very uncertain at these very-high (0.1-100~TeV) and ultra-high ($> 100$~TeV) energies.

\begin{figure*}  
    \centering
   \includegraphics[width=0.99\textwidth]{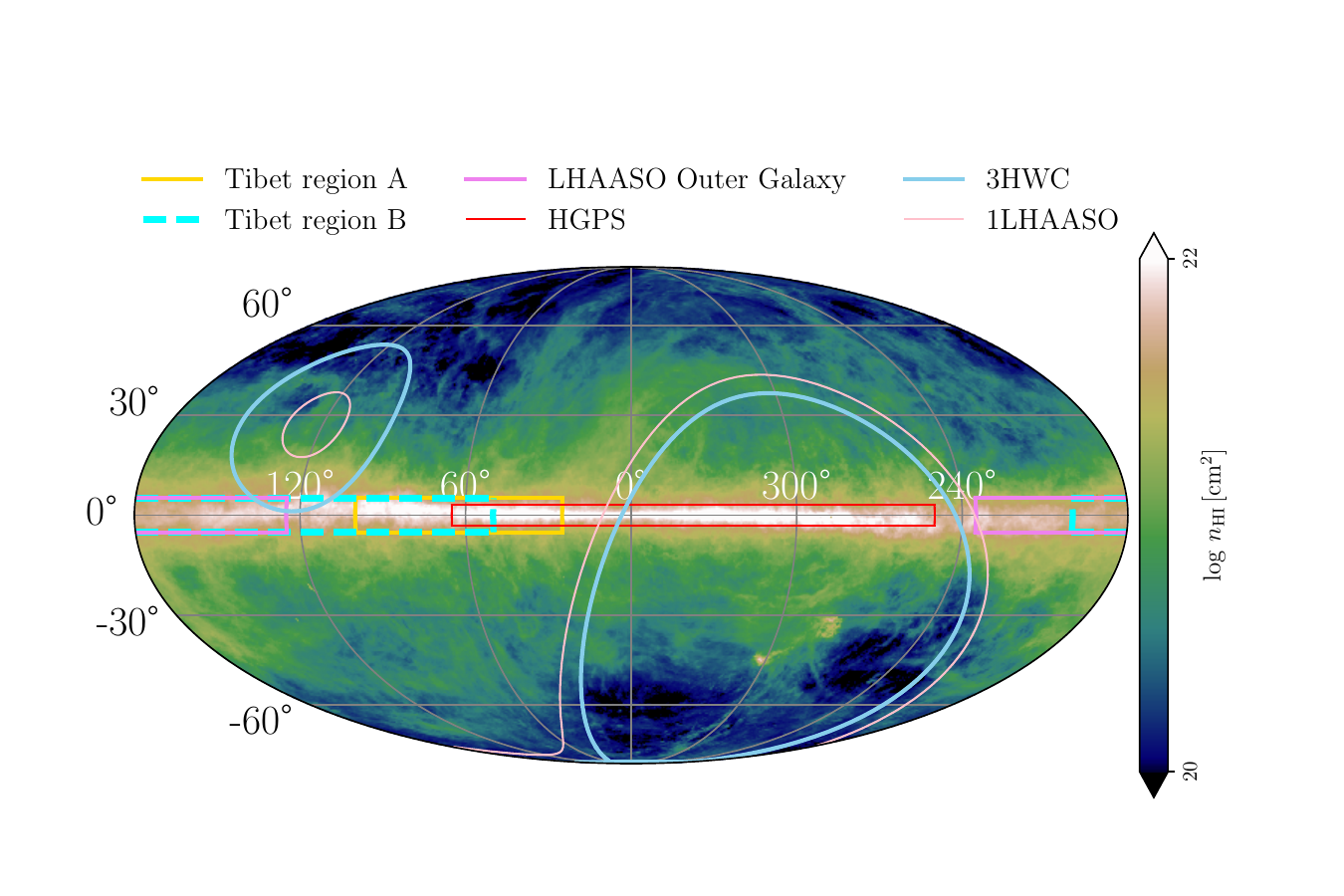}
    \caption{ 
    \label{fig:skymap} Summary of the sky regions observed by various $\gamma$-ray experiments, including H.E.S.S. Telescope for the GP survey (red rectangle; \citealp{HESS:2018pbp}), Tibet AS$\gamma$ Observatory for the GDE observation (yellow rectangle for region A and dashed cyan rectangle for region B; \citealp{TibetDiff}), LHAASO Observatory for the GDE measurement (purple rectangle for outer Galaxy; \citealp{2023arXiv230505372C}), HAWC Observatory for the Third HAWC Catalog of Very-high-energy Gamma-ray Sources (3HWC; sky blue curves \citealp{2020ApJ...905...76A}) and LHAASO for the First LHAASO Catalog of Gamma-Ray Sources (1LHAASO; pink curves; \citealp{2021Natur.594...33C}). {\it Fermi}-LAT and IceCube observe the full sky and are not shown in this plot. Details of the observations are summarized in Table~\ref{tab:table2} and \ref{tab:table1} .  
    For reference, the neutral hydrogen (21~cm) emission from HI 4-PI Survey \citep{2016A&A...594A.116H} is shown with the column density indicated by the color bar. Plot is in Galactic coordinate.
    %For reference, the CO $J= 1\rightarrow0$ intensity measured by the Planck telescope (type 2 line emission map; \citealp{2014A&A...571A..13P}) is shown in colors. 
    }
\end{figure*}

The detection of Galactic neutrinos has been anticipated for decades \citep{1979ApJ...228..919S}.
Whether the Galactic contribution dominates the full-sky neutrino flux was first debated at the time of IceCube's discovery of high-energy cosmic neutrinos \citep{icecubeScience}. Using the multimessenger connection and diffuse TeV $\gamma$-ray data mainly from CASA-MIA and KASKADE, \citet{Ahlers:2013xia} showed that the all-sky neutrino flux mostly originates from extragalactic sources. 
\citet{2021ApJ...919...93F} derived the upper limit on the Galactic neutrino flux based on the GP observation by Tibet AS$\gamma$, and argued that the 100~TeV emission may come from either the GDE or the sum of discrete sources. Lately, the IceCube Collaboration reported evidence for neutrinos from the GP \citep{IceCubeGP}. The observed flux level is consistent with the prediction of \citet{2021ApJ...919...93F}. 

An important task in understanding the GP is to disentangle the contribution of individual sources from the truly diffuse emission. This is crucial to understanding the PeVatrons in the Milky Way and the leptonic contribution to the TeV-PeV $\gamma$-ray sky. 
While detecting individual Galactic neutrino sources would be the ultimate solution to this problem, in this paper we take a first step in understanding the source contribution to the neutrino GDE via a multimessenger approach. Specifically, we constrain the neutrino flux of individual sources using $\gamma$-ray catalogs and compare it to the GDE measured by IceCube or derived from $\gamma$-ray observations. Unlike extragalactic neutrino sources, Galactic neutrino sources are likely optically thin to TeV $\gamma$-rays given their relatively low infrared fluxes. The $\gamma$-ray emission can be made by either electrons or protons and nuclei whereas high-energy neutrinos can only come from the latter. The $\gamma$-ray flux of individual Galactic sources detected by $\gamma$-ray telescopes therefore provide an upper limit on their neutrino emission.   

We describe the TeV-PeV $\gamma$-ray observations of the GP in Section~\ref{sec:TeVg}, including the source catalogs and GDE observations in Section~\ref{sec:cat} and \ref{sec:GDEg}, respectively. By converting the differential $\gamma$-ray flux to neutrino flux assuming that they are simultaneously produced by protons and nuclei, we constrain the high-energy neutrino emission by sources and compare that to the GDE in Section~\ref{sec:nu}. We conclude and discuss the caveats of the work in Section~\ref{sec:dis}.

\section{TeV-PeV Gamma-ray Observations}\label{sec:TeVg}
\begin{figure*}  
    \centering
   \includegraphics[width=0.32\textwidth]{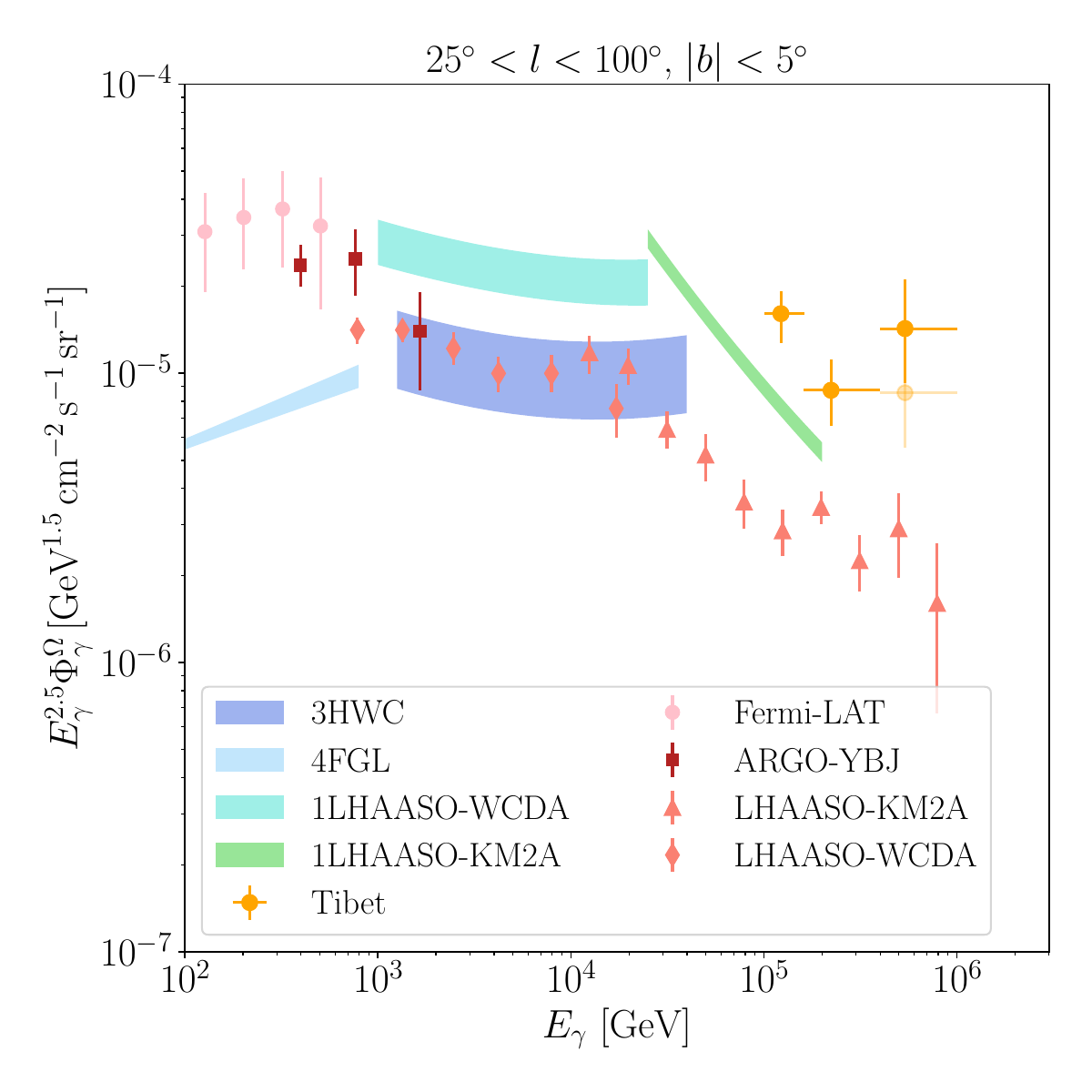}
   \includegraphics[width=0.32\textwidth]{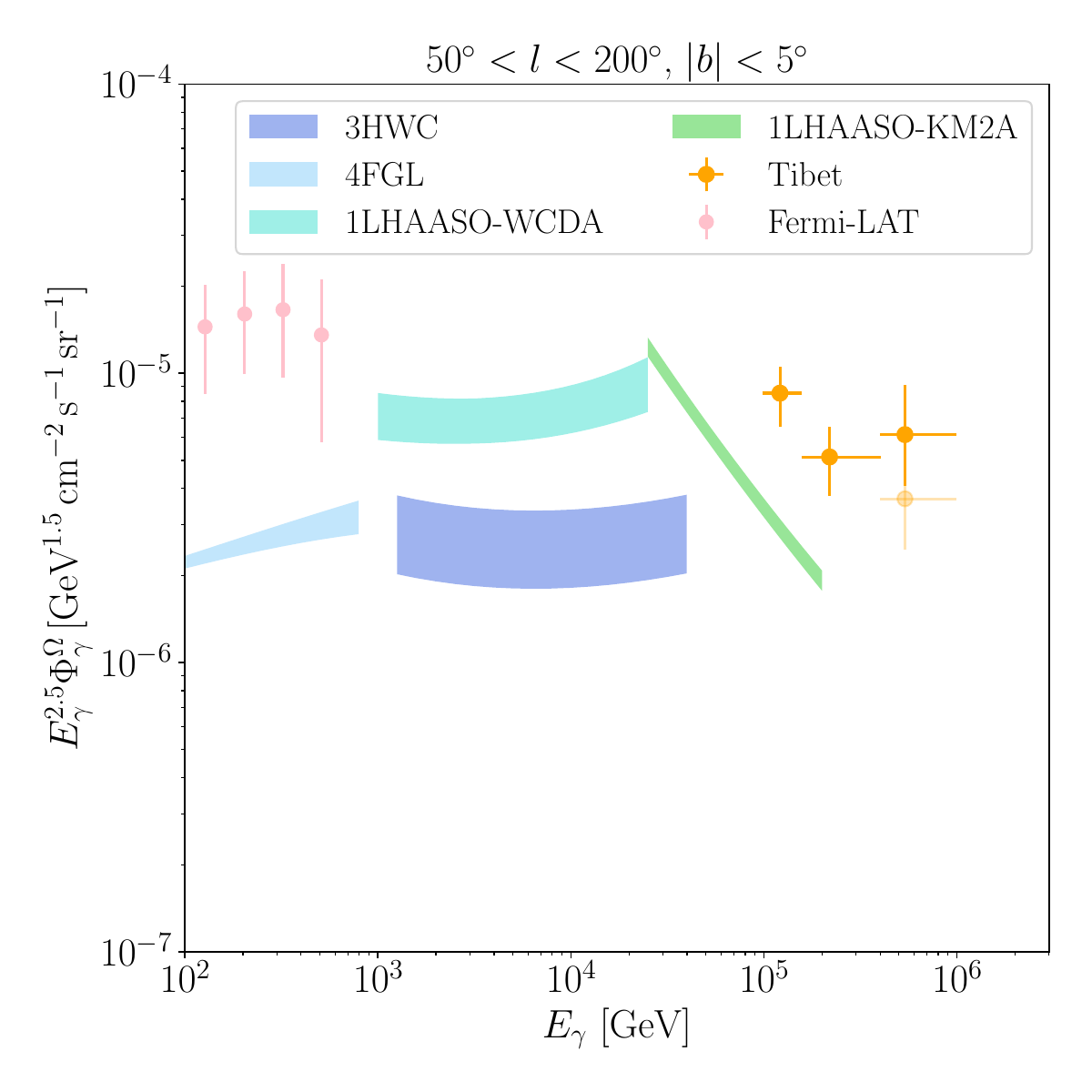}
   \includegraphics[width=0.32\textwidth]{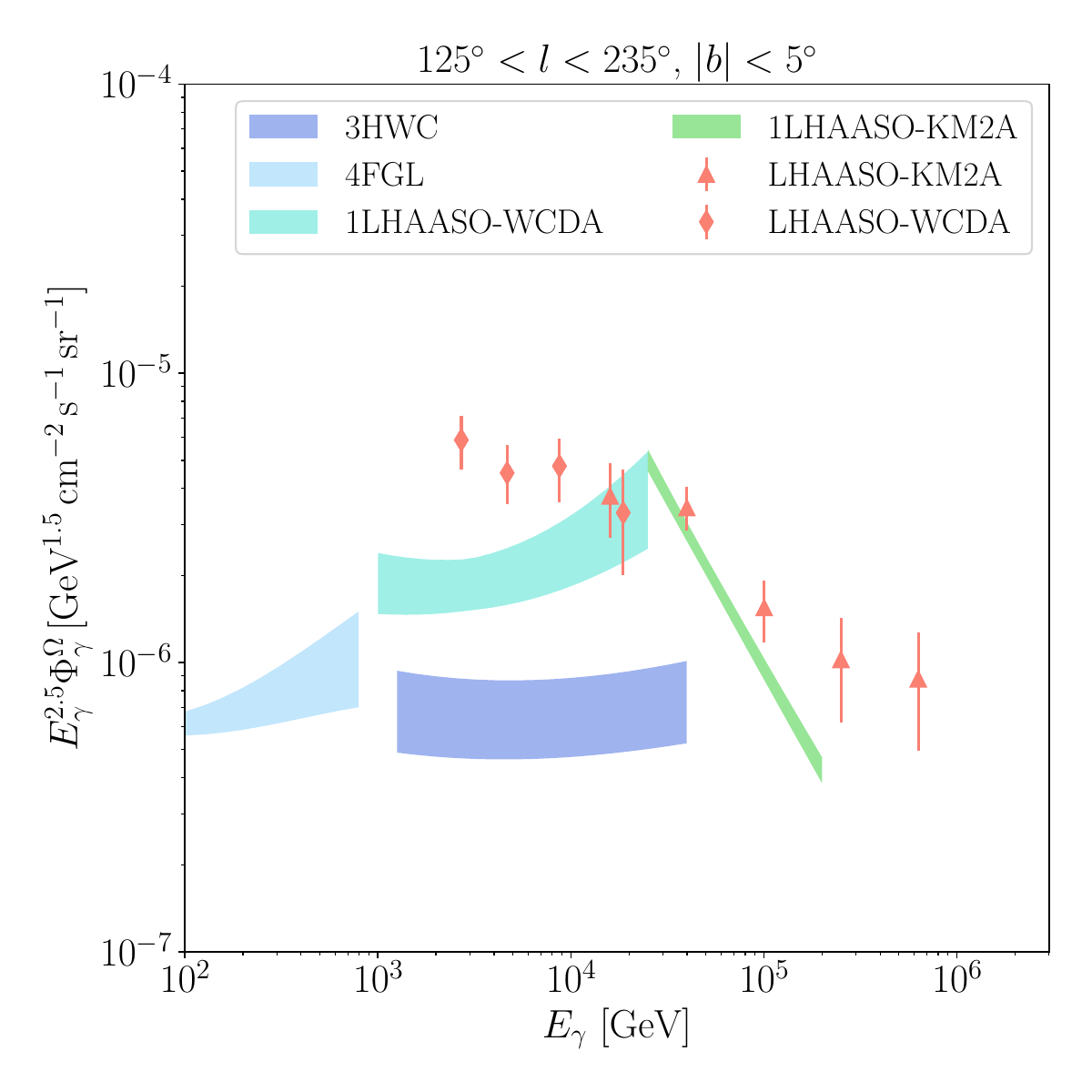}
    \caption{
    \label{fig:gamma_sed} Comparison of intensities of $\gamma$ rays from resolved sources (cool colors) and GDE (warm colors) in three sky regions including (1) Tibet Regions A, (2) Tibet Region B, and (3) LHAASO Outer Galaxy region. The source emissivity is evaluated based on a) 3HWC catalog \citep{2020ApJ...905...76A}, which includes 38, 32, and 10
    sources, b) 4FGL catalog \citep{2022ApJS..260...53A}, which includes  81, 73, and 25 sources, c) 1LHAASO catalog \citep{2023arXiv230517030C}, which includes 37, 34, and 9 sources detected by WCDA, and 40, 37, and 10 sources detected by KM2A
    %c) HAWC's sources above 56~TeV, which includes 6, 2, and 0 sources, and d) LHAASO's sources above 100~TeV, which includes 9, 6, and 0 sources 
    in the three sky regions, respectively. The total source flux is averaged over the solid angle of the corresponding sky regions. For the GDE, the error bars of Tibet AS$\gamma$ observations correspond to $1\,\sigma$ statistical errors and those of the LHAASO flux points correspond to the quadratic sum of the statistical and systematic errors. In the last energy bin of the Tibet AS$\gamma$ GDE flux, the fainter data points indicate the residual intensity after removing the events relevant to Cygnus Cocoon (40\%). In the Tibet Region~A plot, the LHAASO flux points correspond to a similar but larger sky region, the LHAASO inner Galaxy region defined as $15^\circ < l < 125^\circ$ and $|b|<5^\circ$. The {\it Fermi}-LAT data points \citep{2023A&A...672A..58D} correspond to the total flux of the two Tibet regions.}
\end{figure*}

In this section, we describe the $\gamma$-ray catalogs and GDE observations to be used for the deviation of high-energy neutrino fluxes. Figure~\ref{fig:skymap} summarizes the sky regions observed by various experiments. We overlay the neutral hydrogen (HI) emission from the HI 4-PI Survey \citep{2016A&A...594A.116H}, since the pionic GDE is dominated by cosmic-ray interaction with the HI gas.

\subsection{Source Catalogs}\label{sec:cat}
We summarize the sky regions and energy ranges of various $\gamma$-ray source catalogs in Table~\ref{tab:table2} in Appendix~\ref{appendix:table}. Below we describe the usage of each of them. 
 
{\bf HGPS:} 78 sources are reported by the H.E.S.S. Galactic plane survey (HGPS), which is a decade-long observation of the H.E.S.S. telescope with nearly 2700~h of data covering the inner GP \citep{HESS:2018pbp}. One source, HESS~J1943+213, is likely an extragalactic object and is removed from our analysis. For each of the remaining sources, we use the flux at the pivot energy and spectral index reported by the catalog found by assuming a power-law spectral model to derive the differential flux between 1 and 30~TeV. The right end of the energy range is chosen based on the lower limit of the maximum energy of the sources. The 77 Galactic sources include 12 pulsar wind nebulae (PWN), 8 shell-type supernova remnant (SNR), 8 composite SNR (where the emission can come from either the shell or the interior nebula), 3 $\gamma$-ray binaries, and 47 sources without firmly identified associations, including 35 with possible associations in source catalogs and 11 with no associations. We account for a systematic uncertainty of 30\% for the flux. A systematic uncertainty for the spectral index, which is estimated to be an absolute value of 0.2, is not included.

{\bf 3HWC:} 65 sources are reported by the Third HAWC Catalog (3HWC) based on blind searches across HAWC's FOV using 1523 days of data \citep{2020ApJ...905...76A}. Two of them, Mrk~421 and Mrk~501, are extragalactic and removed for the list, yielding a total of 63 Galactic sources. Based on the spectral index and differential flux at a pivot energy of 7~TeV, we calculate the flux of the sources in 3HWC between 1 and 49~TeV. This energy range is within an energy range that contributes to 75\% of the observed significance for most sources. The differential flux of 3HWC is obtained by assuming a pointlike morphology. An extended source may be associated with multiple point sources. The inaccuracy in the source extension barely impact this work since the sum of the flux of point sources reasonably estimates the flux of an extended source. Our calculation includes the systematic uncertainties of the spectral models of the 3HWC sources, which are at the level of 30\%. 

{\bf 1LHAASO:} 90 sources with extension $<2^\circ$ are reported by the first LHAASO catalog (1LHAASO), including 43 sources that are detected at $>4\,\sigma$ above 100~TeV \citep{2023arXiv230517030C}. We exclude the following sources that are likely of extragalactic origin: 1LHAASO~J1104$+$3810, 1LHAASO~J1219$+$2915, 1LHAASO~J1653$+$3943, 1LHAASO~J1727$+$5016, and 1LHAASO~J2346$+$5138. For the remaining sources that are detected, we compute the spectrum following a power law $dN/dE = N_0 (E/E_0)^{-\Gamma}$ between $E_{\rm min}$ and $E_{\rm max}$, with $E_0 = 3$~TeV, $E_{\rm min}=1$~TeV, $E_{\rm max} = 25$~TeV for WCDA and $E_0 = 50$~TeV,  $E_{\rm min}=25$~TeV, $E_{\rm max} = 200$~TeV for KM2A. We include systematic uncertainty of 7\% on KM2A flux and $^{+8\%}_{-24\%}$ on WCDA flux. An absolute uncertanity of 0.02 on spectral index of KM2A measurement is not included. Sources that only have upper limits on flux are not included. 

%{\bf HAWC UHE catalog:} Nine and three sources are detected using HAWC's three-year data above 56~TeV and 100~TeV, respectively \citep{{2020PhRvL.124b1102A}}. All sources are likely of Galactic origin. We derive the differential flux of the sources at 56~TeV using the catalog's integrated flux based on fixed spectral index of $2.7$. The systematic uncertainties of the differential flux is at the level of 15\% at 56~TeV \citep{2019ApJ...881..134A}. 

%{\bf LHAASO UHE catalog:} 12 Galactic sources are found at more than $7\,\sigma$ significance above 100~TeV by LHAASO \citep{2021Natur.594...33C}, which includes the three sources observed by HAWC above 100~TeV. We use the differential flux at 100~TeV reported by the catalog.  

{\bf 4FGL:} Between 50~MeV and 1~TeV, the fourth {\it Fermi} Large Area Telescope catalog (4FGL) reports 6659 sources based 12 years of {\it Fermi}-LAT data \citep{2022ApJS..260...53A}. We count both ``identified" and ``associated" source classes, yielding a total of 539 Galactic sources that can be decomposed into the following groups with corresponding designators: 1) 257 pulsars, including 137 young (`PSR' and `psr') and 120 millisecond pulsars (`MSP'), 2) 20 PWNe (`PWN' and `pwn'), 3) 43 SNRs (`SNR' and `snr') 4) composite SNRs (`spp'), 5) 5 star-forming regions (`SNR' and `sfr'), 6) 26 binaries (`HMB', `hmb', `LMB', `lmb', `BIN', `bin'), 7) 4 novae (`NOV'), 8) 35 globular clusters (`glc'), and 9) Galactic center (`GC'). For each source, we evaluate the differential flux between 0.1 and 1~TeV based on the parameters for the reported {\texttt{SpectrumType}}, which can be a power law, log-parabola, or power law with a super exponential cutoff.  
The errors of the fluxes include systematic uncertainties associated with the detector effective area and Galactic interstellar emission model.

\subsection{Galactic Diffuse Emission}\label{sec:GDEg}
The GDE measurements by various air shower $\gamma$-ray observatories are summarized in Table~\ref{tab:table1} and described below. 

%{\bf CASA-MIA:}  combination of the Chicago Air Shower Array (CASA) and the Michigan Muon Array (MIA) \citep{1998ApJ...493..175B} by comparing muon content on and off the plane. 

{\bf ARGO-YBJ} measured the GDE by subtracting a background map from the event map \citep{2015ApJ...806...20B}. Known sources from the TeVCat were excluded using a $4^\circ\times 4^\circ / \cos(b)$ mask, where $b$ is the latitude. Faint sources were not masked but expected to contribute to $2.5\%$. 

{\bf Tibet AS$\gamma$} detected the GDE at 5.9 $\sigma$ by comparing  the number of $\gamma$-ray-like events from the on region, defined as $|b|<10^\circ$, and the off region, $|b|>20^\circ$. By identifying $\gamma$-ray-like events within $0.5^\circ$ of TeVCat sources, \citet{TibetDiff} concludes that the fractional source contribution to the diffuse component within $|b|<5^\circ$ is $13\%$ above 100~TeV. The events above 398~TeV are likely of a diffuse origin since they neither have accompanying signal at lower energies nor come from directions within $\sim 0.5^\circ$ of known sources. The error bars in the top panels of Figure~\ref{fig:gamma_sed} correspond to $1\,\sigma$ statistical error. In addition, a systematic error of 30\% is expected due to the uncertainty of absolute energy scale \citep{TibetDiff}. 

{\bf LHAASO} detected the GDE from the inner and outer GP at $29.1\,\sigma$ and $12.7\,\sigma$ \citep{2023arXiv230505372C}. Sources detected by KM2A and additional known sources in TeVCat are masked with a Gaussian width that is 2.5 times of the quadratic sum of the point spread function (PSF) of the detector and the source extension. The contribution from remaining resolved sources is estimated to be $<10\%$. The innermost Galactic disk at $15^\circ\lesssim l \lesssim 90^\circ$ and $|b|\lesssim 1.5^\circ$ is mostly masked in the study of \citet{2023arXiv230505372C}, which could have caused an underestimate of the average GDE in that region. \citet{2023arXiv230505372C} found that the flux of the GDE of the inner Galaxy ($15^\circ<l < 125^\circ$ and $|b|\lesssim 5^\circ$) would increase by 61\% when not apply any masking. The GDE flux of the inner Galaxy measured by LHAASO is slightly lower than that of Tibet AS$\gamma$, which could be a result of the more and larger source masks used in LHAASO's analysis. 
%{\ke On the other hand, none of the Tibet AS$\gamma$ diffuse events above 398~TeV comes from the sources detected by LHAASO above 100~TeV, suggesting that a large fraction of Tibet's highest-energy events could be of truly diffusive origin \citep{2023arXiv230916078K}.}
Recently, \citet{Li:2023lm} reports the detection of the diffuse emission from the inner and outer Galactic plane at 27.9~$\sigma$ and 11.9~$\sigma$ significance with the WCDA.

{\bf {\it Fermi}-LAT}: We use the Galactic interstellar emission model (GIEM) for the 4FGL catalog analysis \citep{2022ApJS..260...53A} to evaluate the GDE flux  \footnote{\url{https://fermi.gsfc.nasa.gov/ssc/data/access/lat/BackgroundModels.html}}. We note that the GDE is contributed by both the interstellar emission and unresolved sources, though the fraction of the latter is at percentage level above 10~GeV \citep{2016ApJS..223...26A}. The GIEM is a linear combination of emission components including the $\pi^0$ decay from hadronic cosmic rays interacting with HI gas and molecular hydrogen traced by the CO emission, as well as dark gas, inverse Compton on the interstellar radiation field, and large structures such as the {\it Fermi} Bubbles. The parameters of the model were obtained by fitting to the Pass~8 data. We approximate the flux uncertainty with the systematic uncertainty of the Pass~8 data on the effective area \footnote{\url{https://fermi.gsfc.nasa.gov/ssc/data/analysis/LAT_caveats.html}}, but note that an actual measurement of the GDE could have additional errors associated with the model itself.

\subsection{GDE vs Source Emission in the $\gamma$-Ray Sky}
Figure~\ref{fig:gamma_sed} contrasts the intensites of the $\gamma$-ray emission by resolved sources and the GDE from three sky regions, from inner Galaxy to outer Galaxy: (1) Tibet region A, $25^\circ < l < 100^
\circ$, $|b| < 5^\circ$; (2) Tibet region B, $50^\circ < l < 200^
\circ$, $|b| < 5^\circ$; (3) LHAASO outer Galaxy, $125^\circ < l < 235^
\circ$, $|b| < 5^\circ$. The shaded bands correspond to the sum of sources in the corresponding sky regions. When summing the sources, we add up the flux linearly and the uncertainties in quadrature for error propagation. For the total flux computed using sources from 3HWC, and 1LHAASO catalogs, systematic errors are added with the statistical errors of the flux sum in quadrature, respectively. 

Figure~\ref{fig:gamma_sed} suggests that the GDE is comparable to source emission in the inner Galaxy but may dominate over the source emission in the outer Galaxy. This figure summarizes the flux of resolved sources and GDE in the sky regions observed by Tibet AS$\gamma$ and LHAASO. No scaling factor is applied. Since HGPS has no or partial overlap with these regions (see Figure~\ref{fig:skymap}), the plot does not include the HGPS sources.
 
\section{Neutrino Emission} \label{sec:nu}

Based on the $\gamma$-ray observations in Section~\ref{sec:TeVg}, we derive the upper limit on the Galactic neutrino flux expected from resolved sources and GDE. The connection between $\gamma$-ray and neutrino emission through hadronic processes in the Galaxy is studied in \citet{Ahlers:2013xia,2021ApJ...919...93F} and summarized in Appendix~\ref{appendix:mmConnection}. 
Since none of the TeV $\gamma$-ray experiments covers the full sky, we can only estimate the neutrino emission from the GP using the portion of the plane measured by the $\gamma$-ray detectors, under the assumption that the unobserved region has a similar emissivity distribution as the observed region. Details regarding this deviation are described in Appendix~\ref{appendix:conversionSkyRegions}. The neutrino flux expected from all resolved Galactic $\gamma$-ray sources and the GDE is shown in Figure~\ref{fig:nu_sed} in the Appendix. 

Some classes of $\gamma$-ray sources show clear signatures of leptonic emission. For example, 
%the broadband spectral energy distribution of the Crab nebula can be well described by the synchrotron and inverse Compton emission of relativistic electrons \citep{2020NatAs...4..167H, 2021Sci...373..425L}. 
a systematic study of the population of pulsar wind nebulae (PWNe) in the HGPS catalog suggests that TeV emission by the population can be consistently explained by energetic leptons  \citep{2018A&A...612A...2H}. TeV halos around middle-aged pulsars are a new phenomenon found by air shower detectors \citep{2017Sci...358..911A}. They are much more extended than PWNe, where the electron–positron plasma is confined by the ambient medium. The sizes of TeV halos can usually be explained by the cooling of electrons in the CMB, suggesting that they are also likely of the leptonic origin. 

\begin{figure}[t!]
    \centering
   \includegraphics[width=0.49\textwidth]{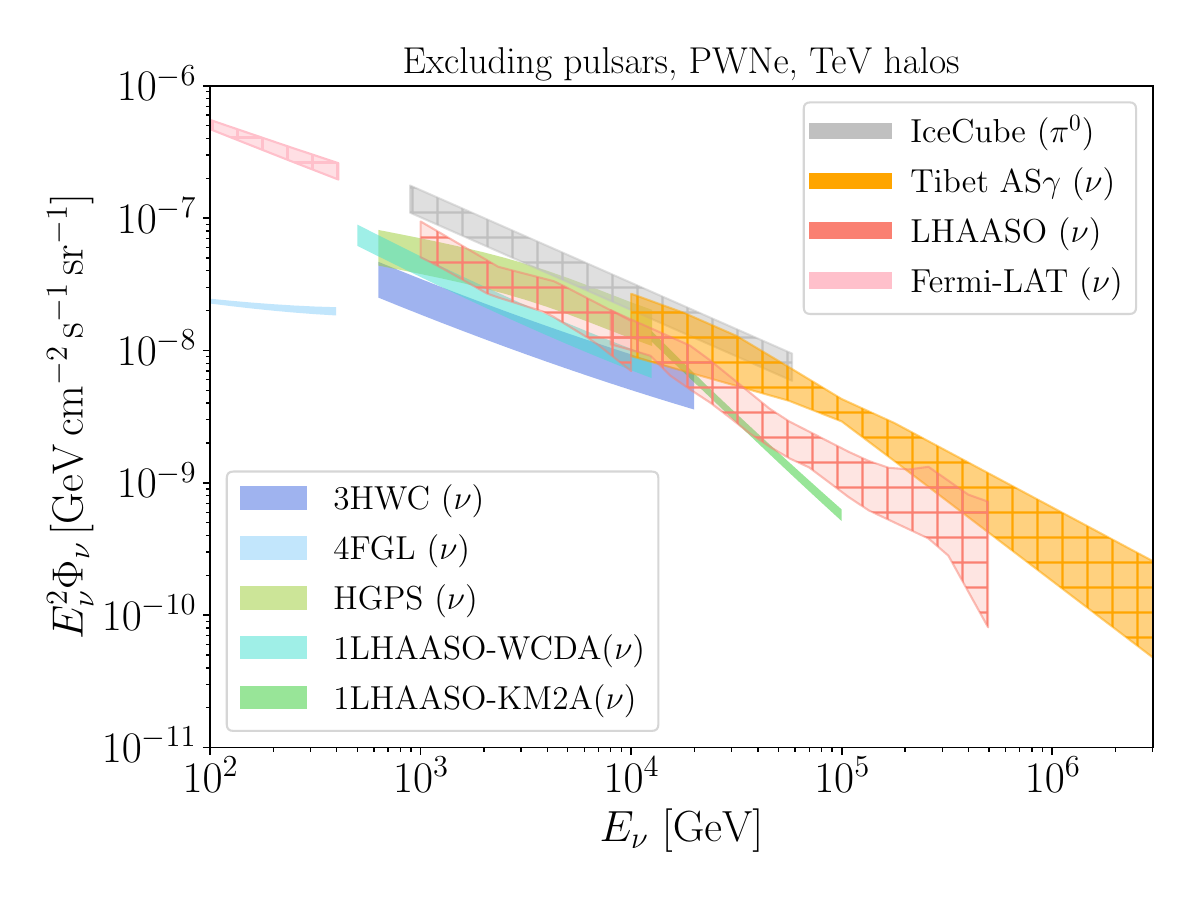}
    \caption{
    \label{fig:nu_sed_noPulsar} All-flavor flux of neutrinos expected from resolved Galactic sources (cool colors, unhatched) and GDE (warm colors, hatched) averaged over the full sky. The source emission is an upper limit based on the assumption that all $\gamma$-ray sources not associated with pulsars are hadronic emitters. The source flux is calculated using the measurements of 227 sources from 4FGL \citep{2022ApJS..260...53A}, 65 sources from HGPS \citep{HESS:2018pbp}, 51 sources from 3HWC \citep{2020ApJ...905...76A}, 36 WCDA sources and 43 KM2A sources from 1LHAASO \citep{2023arXiv230517030C}. The GDE intensity is converted from {\it Fermi}-LAT's Galactic interstellar emission model \citep{2022ApJS..260...53A}, LHAASO \citep{2023arXiv230505372C, Li:2023lm} and Tibet AS$\gamma$'s GDE observations \citep{TibetDiff, 2021ApJ...919...93F}. The hatched grey band is the IceCube measurement of the GP using the $\pi^0$ template \citep{IceCubeGP}.
    }
\end{figure}

Motivated by these facts, we exclude sources in 4FGL and HGPS that are classified as pulsars or PWNe. 
%{\km How many do non-pulsar sources remain for both 4FGL and HGPS?}
We exclude 3HWC sources that are coincident with these TeV halo candidate pulsars (in Table~4 of \citealp{2020ApJ...905...76A}). 
%{\km How many do non-pulsar sources remain for 3HWC?}
For the 1LHAASO catalog, we remove the sources associated with pulsars (in Table~3 of \citealp{2023arXiv230517030C}). 
%{\km How many do non-pulsar sources remain for both WCDA and KM3A?}
In addition, we exclude 1LHAASO~J1831$-$1007u$^*$ and 1LHAASO~J0703$+$1405, which are TeV halo candidates that are removed from the 3HWC. Figure~\ref{fig:nu_sed_noPulsar} presents the neutrino flux of resolved $\gamma$-ray sources that are not associated with pulsars, with the source numbers used for the calculation listed in the caption.

\begin{figure}[t!]
    \centering
   \includegraphics[width=0.49\textwidth]{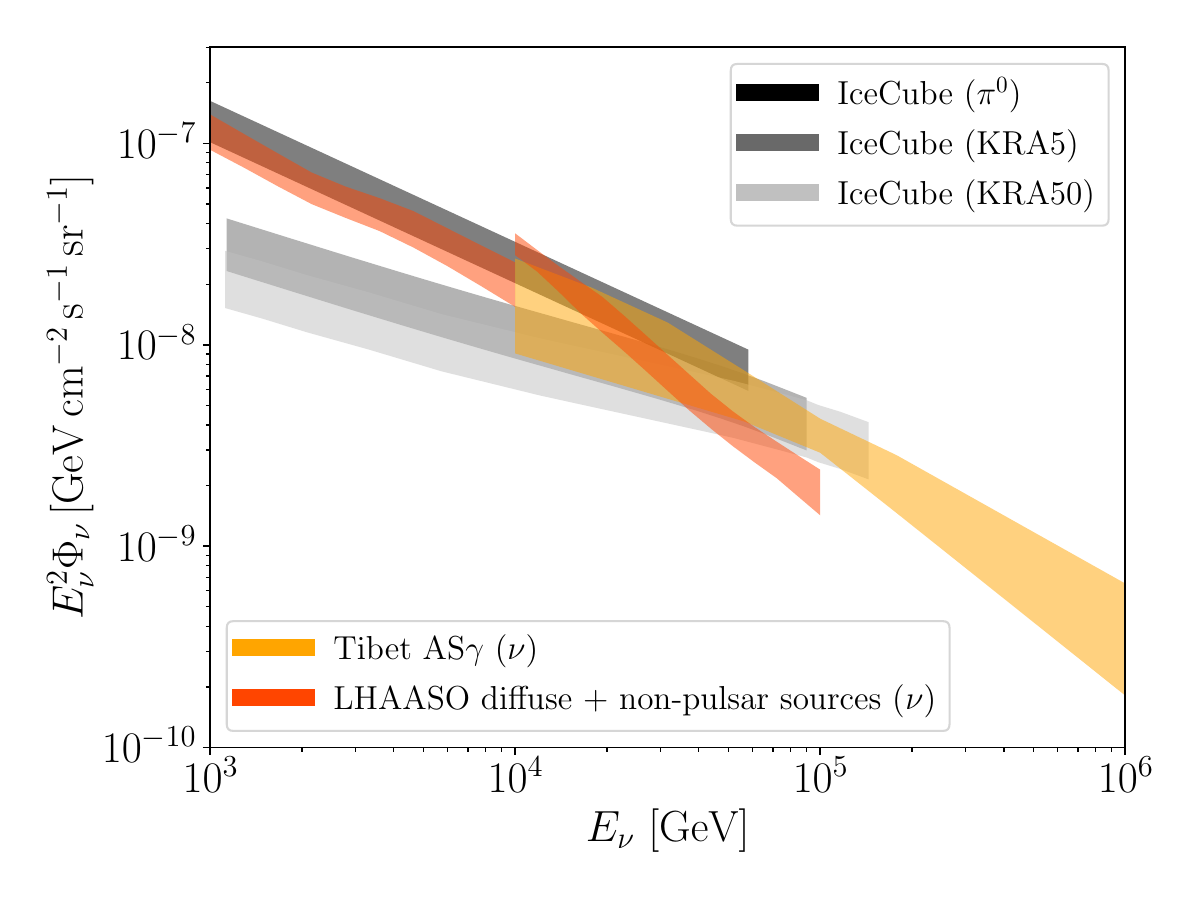}
    \caption{
    \label{fig:nu_sed_addLHAASO} All-flavor, 4$\pi$-averaged flux of neutrinos observed by IceCube with $\pi^0$ (black) and KRA templates (grey and silver) comparing to that derived from the Tibet AS$\gamma$'s GDE observations (orange; \citealp{TibetDiff, 2021ApJ...919...93F}) and LHAASO's GP observations (orange red; \citealp{2023arXiv230517030C, 2023arXiv230505372C, Li:2023lm}). The orange red band sums the flux of the non-pulsar sources and diffuse emission derived from WCDA and KM2A observations in $1-10$~TeV and $10-100$~TeV, respectively. 
    }
\end{figure}

The neutrino GDE flux is derived using the $\gamma$-ray GDE observations listed in Section~\ref{sec:GDEg}. The red band in Figure~\ref{fig:nu_sed_noPulsar} indicates the full-sky GDE derived using the LHAASO observations in both inner and outer Galaxy by assuming that cosmic-ray density follows the SNR distribution described by equation~\ref{eqn:SNR}.
We also overlay the prediction of \citet{2021ApJ...919...93F} based on the Tibet AS$\gamma$ measurement. 
The grey band presents the IceCube measurement of the GDE using the $\pi^0$ template \citep{IceCubeGP}.

Figure~\ref{fig:nu_sed_noPulsar} shows that in an optimistic scenario where all non-pulsar sources are hadronic emitters, the neutrino emission by the sources could be comparable to the GDE at $\sim1-30$~TeV. Above $\sim30-100$~TeV, the neutrino emission from the GP is dominated by the truly diffuse component or unresolved sources that have not been detected by any $\gamma$-ray observations. Given that a significant fraction of the remaining sources are still promising leptonic emitters, such as composite SNRs (e.g., \citealp{2021Univ....7..324C}) and $\gamma$-ray binaries/microquasars (e.g., \citealp{HAWC:2018gwz}), the neutrino emission of the GP is likely dominated by the emission of diffuse cosmic rays and unresolved hadronic sources. 

%Figures~\ref{fig:nu_sed_noPulsar} and \ref{fig:nu_sed} suggest that the spectrum of neutrino emission from the GP due to resolved sources is slightly harder than that arises from the GDE. Measuring both the neutrino spectrum and flux of the GP at $\sim1-30$~TeV can therefore help separate the source and diffuse components. 

%Another IceCube analysis using tracks yielded similar flux \citep{IceCube:2023hou} using the CRINGE model \citep{2023ApJ...949...16S}, the $\pi^0$ model, and the SNR model of  \citet{2021ApJ...919...93F}. \citet{IceCube:2023hou} identifies the CRINGE model as the most preferred model among several diffuse emission templates, though the preference is not statistically significant. It would be important for future Galactic neutrino searches to differentiate the templates. 

The flux of non-pulsar sources and diffuse emission derived from LHAASO observations is summed and shown as the orange red bands in Figures~\ref{fig:nu_sed_addLHAASO}. In particular, we add up the central values of the cyan and green bands, and the red hatched regions in Figure~\ref{fig:nu_sed_noPulsar} for WCDA and KM2A, respectively. We use the band widths as approximate uncertainties and add them in quadrature. Even under the assumption that all non-pulsar sources and diffuse emission are hadronic, their flux sum is comparable to and slightly lower than the IceCube $\pi^0$ flux at $\sim 1-30$~TeV and above $\sim 30$~TeV, respectively. Had the LHAASO diffuse flux been dominated by unresolved leptonic sources such as TeV halos, the remaining diffuse flux would be insufficient to explain the IceCube measurement. Therefore, we conclude that the LHAASO diffuse emission likely comes from hadronic processes especially above $\sim 30$~TeV, either as truly diffuse emission by the cosmic-ray sea or unresolved hadronic sources. The conclusion is subjected to the uncertainty of IceCube measurements especially below 30~TeV as well as the uncertainties arising from our modeling of cosmic-ray distribution described in Appendix~\ref{appendix:conversionSkyRegions}.

\section{Discussion and Conclusions}\label{sec:dis}
We evaluated the GDE and high-energy neutrino flux from astrophysical sources residing in the Milky Way based on the latest $\gamma$-ray observations.  
%Robust upper limits on the TeV-PeV neutrino flux can be placed thanks to the measurements of TeV-PeV $\gamma$-ray fluxes in the past decade. 
Since the TeV-PeV $\gamma$-ray observations are ground-based and cover the partial sky, the maximum flux of neutrino emission from the entire GP is derived based on models of the source distribution in the Galaxy \citep{Ahlers:2013xia,2021ApJ...919...93F}. When calculating the neutrino emission by sources, we removed sources classified as pulsars, PWNe, and TeV halos which are promising leptonic sources. Our main conclusions are summarized as follows.

% Resolved pulsar sources/PWNe dominate over the diffuse or unresolved contribution
% Resolved non-pulsar gamma-ray sources that can be hadronic sources give a contribution to the contribution from diffuse/unresolved sources.
% The sum of resolved+diffuse/unresolved matches the IceCube flux (with pi^0 template).
% Interpretation 1: The IceCube flux may come from ~50% of the diffuse/unresolved and ~50% from the resolved contribution 
%Interpretation 2: The tmeplates are wrong (e.g., KRA), and then the ICeCube flux may predominantly come from the diffuse/unresolved.
% For >100 TeV, the resolved non-pulsar contribution seem to be less than the diffuse/unresolved, so there is a possibility that the contribution from unresolved super-PeVatrons can be more important (This is because the truly diffuse component typically declines above 100 TeV due to the knee.)

\begin{itemize} 
  \item The neutrino contribution from resolved $\gamma$-ray sources, excluding those associated with pulsars, is smaller than the IceCube Galactic neutrino flux measured with the $\pi^0$ template by a factor of $\sim$2, suggesting that the neutrino emission could be dominantly produced by diffuse cosmic rays or sources unresolved by $\gamma$-ray facilities.  % in the neutrino sky although the current data are still subject to significant uncertainties. %This is especially the case if the GP flux derived for the $\pi^0$ template \citep{IceCubeGP} is correct.
  
  \item At $\sim 1-30$~TeV, the sum of resolved non-pulsar sources and the LHAASO diffuse emission is comparable to the IceCube $\pi^0$ flux, when assuming that the $\gamma$-ray emission of these components is 100\% hadronic. This indicates that the LHAASO diffuse $\gamma$-ray emission could not be dominated by unresolved leptonic sources such as TeV halos. The above two conclusions are weaker when comparing to the IceCube flux obtained with KRA templates. 
  %km: unlikely should be used for the tension with 2 or 3 sigma or larger.... 

  %{\km \item  At $\sim 30-100$~TeV, the neutrino flux measured by IceCube is unlikely to be smaller than that derived from the Tibet AS$\gamma$ GDE \citep{2021ApJ...919...93F}, suggesting a hadronic origin of the GP $\gamma$-ray emission.} 

   \item  At $\sim 30-100$~TeV, the neutrino flux measured by IceCube is comparable to or higher than that derived from the Tibet AS$\gamma$ GDE \citep{2021ApJ...919...93F}, suggesting a hadronic origin of the GP $\gamma$-ray emission.

 % \item Above $\sim 100$~TeV, the resolved non-pulsar $\gamma$-ray flux is smaller than the LHAASO diffuse $\gamma$-ray flux, which consists of the GDE and unresolved $\gamma$-ray sources. Because the Galactic neutrino flux observed in IceCube already matches the neutrino flux converted from the Tibet AS$\gamma$ data \citep{2021ApJ...919...93F} at $\sim10-30$~TeV, the hadronic origin would be favored for these non-pulsar $\gamma$-ray sources. 
  \item Above $\sim100$~TeV, the GDE is expected to decline due to the cosmic-ray knee.  If the GP neutrino and $\gamma$-ray spectra extend to higher-energy without a break, then it would be natural to expect contribution from super-PeVatrons such as hypernova remnants and super-bubbles \citep{Ahlers:2013xia,Zhang:2019sio, 2021NatAs.tmp...50A} 
\end{itemize}
%We assumed that the GDE above 10~TeV is dominated by diffuse emission, although it could also come from unresolved sources. The unresolved Galactic sources are estimated to contribute to the GeV-TeV Galactic interstellar emission at the level of 3\% \citep{2015ApJS..218...23A}. If TeV-PeV sources follow a similar luminosity distribution as the lower-energy $\gamma$-ray sources, unresolved sources would barely impact our conclusion. On the other hand, 

The identification and measurement of Galactic neutrino or $\gamma$-ray sources involve a separation of the GDE component. A small fraction of the source flux could arise from the GDE and the isotropic emission \citep{2023arXiv230517030C}. This would further lower the source contribution and support our conclusion. 

We assumed that $\gamma$-ray emission of pulsars, PWNe, and TeV halos mostly come from relativistic electrons and positrons. High-energy neutrinos could be emitted by fast-spinning newborn pulsars, although the birth rate of such sources in the local Universe is relatively low \citep{Bednarek:1997cn,Murase:2009pg,Fang:2014qva}. 

Our results confirmed the previous findings that the Galactic contribution, whether it originates from truely diffuse emission or sources, should be subdominant in the all-sky neutrino flux in the 10~TeV -- 1~PeV
range \citep{Ahlers:2013xia, Murase:2015xka, 2018A&A...615A.168P, 2021ApJ...919...93F}. Although our conclusion is not directly applied to quasi-isotropic emission, this has also been constrained by not only {\it Fermi}-LAT but also TeV-PeV $\gamma$-ray observations \citep{Murase:2013rfa,Ahlers:2013xia,Murase:2015xka}. 

Upcoming neutrino telescopes such as KM3Net, Baikal-GVD and IceCube-Gen2 \citep{2020arXiv200804323T} may resolve individual Galactic sources and disentangle the source emission and GDE. 
Future air shower $\gamma$-ray experiments in the Southern hemisphere such as the Southern Wide-field Gamma-ray Observatory \citep{2019arXiv190208429A} are also crucial to understanding the emission of the entire GP. 

\begin{acknowledgments}
%We thank xxx for helpful comments on the manuscript. 
The work of K.F. is supported by the Office of the Vice Chancellor for Research and Graduate Education at the University of Wisconsin-Madison with funding from the Wisconsin Alumni Research Foundation. K.F. acknowledges support from National Science Foundation (PHY-2110821, PHY-2238916) and from NASA through the Fermi Guest Investigator Program (80NSSC22K1584). 
The work of K.M. is supported by the NSF grants No.~AST-1908689, No.~AST-2108466 and No.~AST-2108467, and KAKENHI No.~20H01901 and No.~20H05852.
\end{acknowledgments}

\bibliography{references,kmurase}

%=====================================
%              APPENDIX
%=====================================

\appendix

\section{Table Summary of Gamma-ray observations of the Galaxy}\label{appendix:table}
Table~\ref{tab:table2} and \ref{tab:table1} summarize the $\gamma$-ray observations of Galactic sources and GDE, respectively. 

\citet{2023A&A...672A..58D} compared $\gamma$-ray emission models to the {\it Fermi}-LAT data from the two sky regions observed by Tibet AS$\gamma$. They conclude that the total flux is dominated by the $\pi^0$ decay of the diffuse cosmic rays at 100-300~GeV, with $<10\%$ contributed by resolved and unresolved sources, inverse Compton and bremsstrahlung radiation from cosmic-ray electrons, and the isotropic $\gamma$-ray background. We therefore use the total flux of the {\it Fermi}-LAT data from \citet{2023A&A...672A..58D} as an approximate of the GDE flux in these two regions (the left and middle plots in Figure~\ref{fig:gamma_sed}). 

%%%%%%%%%%%%%%%%%
%%%% TABLE 1 %%%%
%%%%%%%%%%%%%%%%%
\begin{deluxetable*}{lccccccc}
\centering
\tablecaption{Summary of sky regions observed by $\gamma$-ray experiments for source catalogs. 
\label{tab:table2} 
}
\tablecolumns{5}
\tablewidth{0pt}
\tablehead{\colhead{Experiment} & \colhead{Catalog}	& \multicolumn{2}{c}{Sky regions} & Energy [TeV] & \colhead{Reference}  } 
\startdata
{\it Fermi}-LAT    & 4FGL  &    all-sky  &   &  $10^{-4}-1$ &\citet{2022ApJS..260...53A}  \\
\hline
H.E.S.S.     & Galactic Plane survey   &   $250^\circ \leq l \leq 65^\circ$ & $|b|\leq 3^\circ$    &  $>1$  &  \citet{HESS:2018pbp}  \\
\hline
HAWC    &  3HWC  &   $-26^\circ < \delta  < 64^\circ$ &   $0^\circ < \alpha < 360^\circ$   &  $7$  &  \citet{2020ApJ...905...76A} \\
%   &  sources above 56~TeV &  &   &  $>56$, $>100$  &  \citet{2020PhRvL.124b1102A} \\
\hline
%LHAASO   & sources above 100~TeV  &   $-15^\circ < \delta  < 75^\circ$ &   $0^\circ < \alpha < 360^\circ$    &     $>100$ &\citet{2021Natur.594...33C} \\
LHAASO   & 1LHAASO  &   $-20^\circ < \delta  < 80^\circ$ &   $0^\circ < \alpha < 360^\circ$    &     $>1$ &\citet{2023arXiv230517030C} \\
\hline
\enddata
%\tablenotetext{{\star}}{}
\end{deluxetable*}

%%%%%%%%%%%%%%%%%
%%%% TABLE 2 %%%%
%%%%%%%%%%%%%%%%%
\begin{deluxetable*}{lccccccc}
\centering
\tablecaption{Summary of GDE measurements by $\gamma$-ray experiments. 
\label{tab:table1} 
}
\tablecolumns{6}
\tablewidth{0pt}
\tablehead{\colhead{Experiment} & \colhead{Observation}	& \multicolumn{2}{c}{Sky regions} & Energy [TeV] &  \colhead{Reference}  } 
\startdata
ARGO-YBJ       &  GDE region A  &   $25^\circ \leq l \leq 100^\circ$ & $|b|\leq 5^\circ$ & $0.35-2$     &  \citet{2015ApJ...806...20B} \\
\hline
Tibet AS$\gamma$     &  GDE region A   &   $25^\circ \leq l \leq 100^\circ$ & $|b|\leq 5^\circ$  & $100-1000$   &  \citet{TibetDiff} \\
     &  GDE region B   &   $50^\circ \leq l \leq 200^\circ$ & $|b|\leq 5^\circ$   & $100-1000$    &  \citet{TibetDiff} \\
\hline 
LHAASO  &  GDE  inner Galaxy &   $15^\circ < l < 125^\circ$ & $|b| \leq 5^\circ$  &  $10-1000$   &  \citet{2023arXiv230505372C, Li:2023lm} \\
     &  GDE  outer Galaxy &   $125^\circ < l < 235^\circ$ & $|b| \leq 5^\circ$  &  $10-1000$ &  \citet{2023arXiv230505372C, Li:2023lm} \\
\hline
\enddata
%\tablenotetext{{\star}}{}
\end{deluxetable*}

\section{Multimessenger Connection}\label{appendix:mmConnection}
%conversion between regions, IR absorption, LOS scaling based on emissivity models 
As in \citet{2021ApJ...919...93F}, we derive the upper limit on the neutrino flux of a sky region from the $\gamma$-ray measurements through the following relation: 
\begin{eqnarray}
\label{eqn:Fnu_Fg}
&& E_\nu^2 F_\nu^\Omega \approx \frac{3}{2} \left.\left(E_\gamma^2 F_\gamma^\Omega\right)\right|_{E_\gamma = 2E_\nu} \\ \nonumber
&\times&\frac{\int ds \int\cos{b}\, db\int dl \, n_s (s, b, l)}{ \int ds \int\cos{b} db \int dl \, n_s P_{\gamma,\rm surv}(E_\gamma = 2E_\nu, s, b, l)},
\end{eqnarray}
where $F_\nu^\Omega$ and $F_\gamma^\Omega$ are the all-flavor neutrino flux and $\gamma$-ray flux produced by hadronic cosmic rays from a sky region, either as GDE or source emission. The factor to the right hand side of the equation scales the emissivity of the sky regions by accounting for the attenuation of $\gamma$-rays due to propagation in the ISM. In particular, $P_{\gamma,\rm surv}$ is the probability for a photon to survive from the pair production along a line-of-sight $s$ in the direction of Galactic longitude $l$ and latitude $b$,  
\begin{equation}
P_{\gamma, \rm surv}(E_\gamma, \vec{x}_0,  \vec{x}_{\rm ob}) = \exp\left(-\tau_{\gamma\gamma}(E_\gamma, \vec{x}_0,\vec{x}_{\rm ob})\right), 
\end{equation}
and $\tau_{\gamma\gamma}$ is the optical depth to a photon with energy $E_\gamma$ when traveling from its initial position $\vec{x}_0$ to the observer at $\vec{x}_{\rm ob}$ computed using the CMB and the interstellar radiation field model of \citet{Vernetto_Lipari16}. 

The integrant $n_s$ is the number density of $\gamma$-ray and neutrino emitters at position $(s, b, l)$. In the case of source emission, it is equivalent to the source density, $n_s = n_{\rm CR}$. In the case of diffuse emission, it is proportional to the product of the cosmic ray ($n_{\rm CR}$) and gas and molecular densities $n_N$,  $n_s \propto n_{\rm CR}n_N$. We approximate $n_N$ with the HI gas density based on the model of \citet{Nakanishi:2003eb, Evoli:2016xgn}. For the diffuse emission calculation, we have assumed that the contribution of unresolved sources is so small that the emissivity scales to $n_{\rm CR}n_N$ instead of $n_{\rm CR}$. We have also assumed that the cosmic-ray density is proportional to the source density, although the former could be smoother than the latter due to the effect of cosmic-ray diffusion in the Galactic magnetic field.

%For $E_\gamma \ll 100$~TeV, $P_{\gamma, \rm surv} \approx 1$, the scaling factor at the right hand side of equation~\ref{eqn:Fnu_Fg} is one and 
When the effective attenuation factor at the right hand side of equation~\ref{eqn:Fnu_Fg} is 1, the equation returns to the usual form of equation~2 of \citet{Ahlers:2013xia}.

\section{Conversion among different sky regions}
\label{appendix:conversionSkyRegions}
We derive the neutrino emission of the entire GP from partial-sky observations under the assumption that the unobserved region has a similar emissivity distribution as the observed region. This is done using equation~\ref{eqn:Fnu_Fg} but integrating over different sky regions for neutrinos, $\Omega_\nu$, and $\gamma$-ray, $\Omega_\gamma$. 

When converting source emission, we take $\Omega_\nu = 4\pi$ and $\Omega_\gamma$ of various source catalogs and assume that sources follow the spatial distribution of supernova remnants (SNR). 

\begin{equation}\label{eqn:SNR}
n_{\rm CR}\propto \left(\frac{r}{R_\odot}\right)^\zeta\exp{\left[-\eta\left(\frac{r-R_\odot}{R_\odot}\right)-\frac{|z|}{z_g}\right]}.
\end{equation}
where $R_\odot=8.5$~kpc is the solar distance from the GC and  the following parameter values are adopted,   $\zeta=1.09$, $\eta = 3.87$ \citep{2015MNRAS.454.1517G} and $z_g=0.083$~kpc \citep{Steppa:2020qwe}. 

%or is uniform within the plane (as defined by equation~8 of \citealp{2021ApJ...919...93F}). % and scaled by the $pp$ optical depth \citep{Ahlers:2013xia}. 

\section{Neutrino emission from all sources}
\begin{figure} [t]
    \centering
   \includegraphics[width=0.49\textwidth]{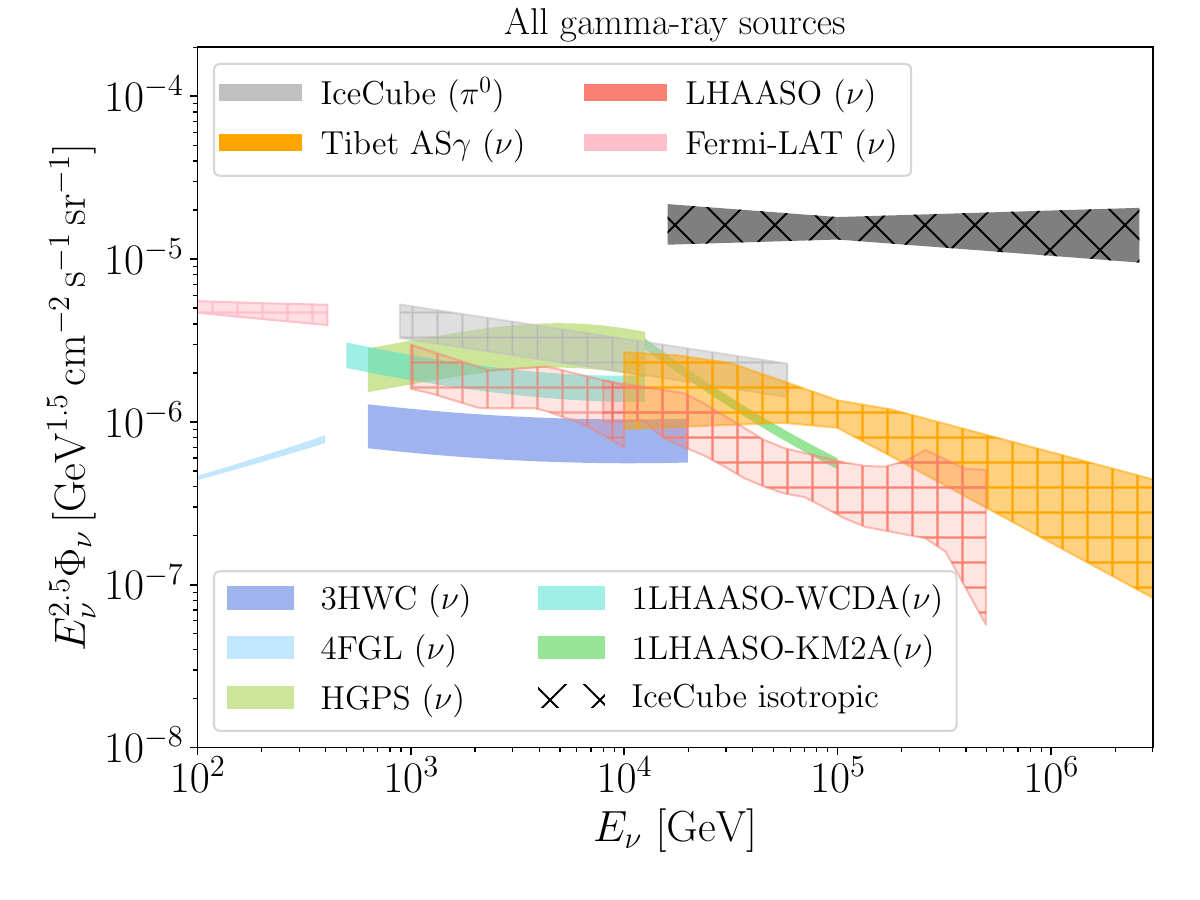}
    \caption{
    \label{fig:nu_sed} Same as Figure~\ref{fig:nu_sed_noPulsar} but in $E_\nu^{2.5}\Phi_\nu$ and including all $\gamma$-ray sources. The source emission is evaluated based 63 sources from 3HWC, 539 sources from 4FGL, 64 sources detected by WCDA and 75 sources detected by KM2A from 1LHAASO, and 77 sources from HGPS. The IceCube isotropic flux is overlaid (in black hatched band; \citealp{IceCube:2020acn}). 
    }
\end{figure}

Figure~\ref{fig:nu_sed} contrasts the fluxes of the neutrinos expected from all resolved sources in the Galaxy and the GDE. Since the conversion is based on an optimistic assumption that all $\gamma$-ray emission is produced by cosmic-ray protons and nuclei in astrophysical sources, the resulted fluxes should be treated as upper limits. 

Around $10$~TeV, the source flux derived from the HGPS catalog is a few times higher than that from the 1LHAASO and 3HWC catalogs. The sensitivities of the HGPS and 3HWC are comparable \citep{HESS:2018pbp, 2020ApJ...905...76A}. The comparison of the GP observed by H.E.S.S. and HAWC at $10^\circ < l < 60^\circ$ leads to similar integrated fluxes above 1~TeV
\citep{2021ApJ...917....6A}. 
As the HGPS covers only a small range of latitudes ($|b| < 3^\circ$), the relatively high neutrino flux derived from the HGPS catalog is probably due to the fact that the SNR model (equation~\ref{eqn:SNR}) used for the conversion does not sufficiently describe the clustering of $\gamma$-ray sources in the inner Galaxy. Furthermore, more than half of the HGPS region is in the Southern sky, which is not accessible to LHAASO and HAWC (see Figure~\ref{fig:skymap}). Future air shower $\gamma$-ray facilities in the Southern sky are needed to fully understand the difference. 

%\section{Comparison with theoretical models}
\begin{figure}[t!]
    \centering
   \includegraphics[width=0.49\textwidth]{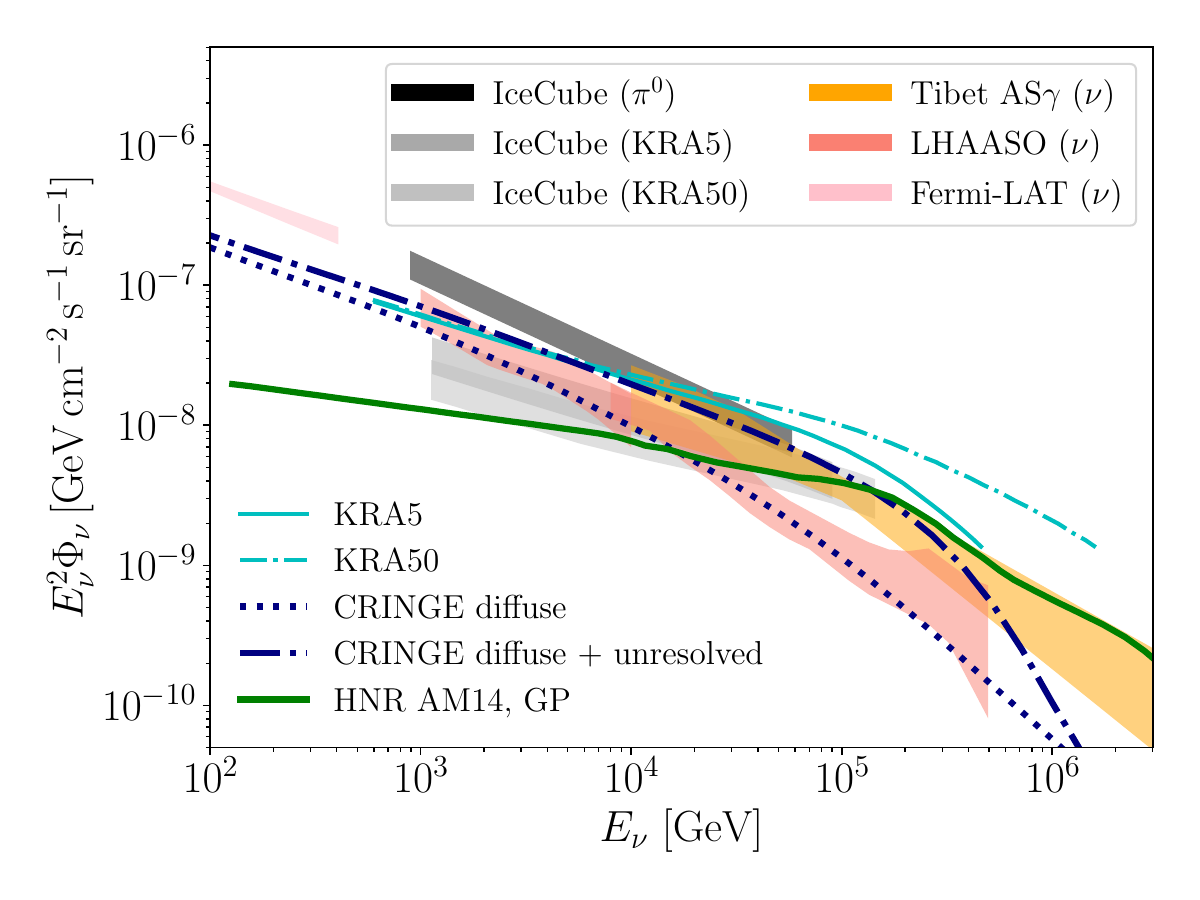}
    \caption{
    \label{fig:nu_sed_Model} Measured and derived all-flavor neutrino flux from GDE averaged over the full sky (warm colors, unhatched) comparing with models, including the KRA models \citep{2015ApJ...815L..25G}, the CRINGE models \citep{2023ApJ...949...16S}, and the HNR model \citep{Ahlers:2013xia}. 
    }
\end{figure}

In Figure~\ref{fig:nu_sed_noPulsar} and \ref{fig:nu_sed}, the GP flux corresponds to the IceCube measurement using the shower data and the $\pi^0$ template \citep{IceCubeGP}. Another IceCube analysis using tracks \citep{IceCube:2023hou} yields a similar flux with $\pi^0$ and CRINGE \citep{2023ApJ...949...16S} templates. Comparing to the $\pi^0$ and CRINGE flux, the GP flux derived from the shower data \citep{IceCubeGP} with the KRA templates \citep{2015ApJ...815L..25G} is similar above $\sim 10$~TeV but 3-5 times lower around 1~TeV. Figure~\ref{fig:nu_sed_addLHAASO} and \ref{fig:nu_sed_Model} present the KRA flux from \citet{IceCubeGP}. The KRA flux from showers is about twice lower than that obtained from tracks \citep{IceCube:2023hou}.

Figure~\ref{fig:nu_sed_Model} further compares theoretical models with the derived and measured neutrino GDE. It is intriguing that the sum of unresolved HNRs \citep{Ahlers:2013xia}, with the gamma-ray flux converted to the neutrino flux following our method, can match the maximum Galactic neutrino flux derived from the Tibet AS$\gamma$ data \citep{2021ApJ...919...93F}.

\end{document}